\documentclass[12pt]{article}

\usepackage{amssymb,amsmath}
\usepackage{graphics}
\usepackage{epsfig}
\usepackage{amsfonts}

\usepackage{psfrag}

\usepackage{amscd}

\usepackage{epsfig,amsfonts,amssymb}
\usepackage{amsmath}
\usepackage{amssymb}
\usepackage{txfonts} 
\usepackage{mathrsfs}
\usepackage{cite}
\begin{document}
\begin{center}
{\Large  On the universal hydrodynamics of strongly coupled CFTs with gravity duals}
\end{center}
\centerline{\large \rm Rajesh Kumar Gupta, Ayan Mukhopadhyay}

\centerline{\large \it Harish-Chandra Research Institute}

\centerline{\large \it  Chhatnag Road, Jhusi,
Allahabad 211019, INDIA}
\vspace*{1.0ex}

\centerline{E-mail:
rajesh@mri.ernet.in, ayan@mri.ernet.in}

\vspace*{5.0ex}
\begin{abstract}
It is known that the solutions of pure classical 5D gravity with $AdS_5$ asymptotics can describe strongly coupled large N dynamics in a universal sector of 4D conformal gauge theories. We show that when the boundary metric is flat we can uniquely specify the solution by the boundary stress tensor. We also show that in the Fefferman-Graham coordinates all these solutions have an integer Taylor series expansion in the radial coordinate (i.e. no $log$ terms). Specifying an arbitrary stress tensor can lead to two types of pathologies, it can either destroy the asymptotic AdS boundary condition or it can produce naked singularities. We show that when solutions have no net angular momentum, all hydrodynamic stress tensors preserve the asymptotic AdS boundary condition, though they may produce naked singularities. We construct solutions corresponding to arbitrary hydrodynamic stress tensors in Fefferman-Graham coordinates using a derivative expansion. In contrast to Eddington-Finkelstein coordinates here the constraint equations simplify and at each order it is manifestly Lorentz covariant. The regularity analysis, becomes more elaborate, but we can show that there is a unique hydrodynamic stress tensor which gives us solutions free of naked singularities. In the process we write down explicit first order solutions in both Fefferman-Graham and Eddington-Finkelstein coordinates for hydrodynamic stress tensors with arbitrary $\eta/s$. Our solutions can describe arbitrary (slowly varying) velocity configurations. We point out some field-theoretic implications of our general results.
\end{abstract}
\section{Introduction}
In one of the major developments of late 20-th century physics, it has been shown that many strongly coupled conformal 4D gauge theories at large N can be solved by using a classical theory of gravity in ten dimensional spacetime with $AdS_5 \times X$ asymptotics\cite{9711200,9802150,9802109}. X is a compact Sasaki-Einstein manifold and is related to the R symmetry of the theory if the gauge theory is supersymmetric. In the classical theory of gravity the dynamics of the metric will be described by Einstein's equation sourced by a matter energy-momentum tensor. The matter content of the theory of gravity will depend on the presumed dual gauge theory. By the gauge/gravity duality any smooth solution of the equations of motion of the classical theory of gravity is dual to an on-shell state in the conformal gauge theory and encodes all the dynamics of the strongly coupled CFT state in a precise way \cite{9802150,0209067}. 

There is, however, always a sector of the theory where the dynamics is universal. This is because any two-derivative theory of classical gravity which has $AdS_5 \times X$ as a solution always admits a consistent truncation to five dimensional Einstein's equation with a negative cosmological constant. For instance, we can set all scalar fields arising from Kaluza Klein excitations on X and other sources to values that minimise the potential and turn off all other matter fields. 

Using AdS/CFT correspondence, now we can define the universal sector of all strongly coupled (large N) conformal field theories with gravity duals as follows. This sector by definition is the dual of pure $5$-dimensional gravity with asymptotic AdS boundary condition. A state in this universal sector will be dual to a smooth solution of Einstein's equation with negative cosmological constant. At finite temperature also, this correspondence works, but now the solutions of pure classical gravity are required to be free of naked singularities.\footnote{This universal sector is different from what in the context of calculating the tachyon vacuum in string field theory is also called the universal sector of 2D CFTs. In the latter case, it is defined to be the set of states generated by the action of Virasoro generators on the vacuum \cite{9911116}. However these states cannot be uniquely specified just by the vev of stress tensor alone whereas all solutions of pure gravity can be uniquely specified by the boundary stress tensor. So even for 2D CFTs our universal sector (which can be defined to be the dual of pure 3D gravity with negative cosmological constant) is different from the other definition.}
 
In the first part of the paper we will argue that all solutions of pure classical gravity in the universal sector with $AdS_5$ asymptotics are uniquely determined by the boundary stress tensor when the boundary metric is flat. The $AdS_5$ asymptotics always requires a choice of a boundary conformal structure which means that the induced metric on the surface at infinity has a double pole in the radial coordinate and its residue can only be fixed upto conformal transformations in the boundary coordinates. We say that the boundary metric is flat when we choose the boundary conformal structure to be that of flat space. In the gauge/gravity dictionary it translates into the dual CFT living in flat space. So our result implies that in the universal sector the strongly coupled dynamics of the CFT state at large N is specified once the conservation of the expectation value of the traceless stress tensor is satisfied. From the field-theoretic point of view, this is a surprising simplification of the dynamics.

To establish our claim we will use a theorem due to Fefferman and Graham \cite{Fefferman}, which states that for any solution of Einstein's equations with $AdS$ asymptotics we can always use a certain coordinate system within a finite distance from the boundary. Skenderis and others \cite{0209067,0002230} have shown that this Fefferman-Graham coordinate system also captures the physics of the CFT nicely, in particular, one can read off the expectation values of various operators in the dual CFT state and also the Weyl anomaly directly from the metric in this coordinate system. We will use some characteristics of the CFT to argue that when the boundary metric is flat the metric in Fefferman-Graham coordinates should have a simple integer Taylor series expansion in the radial coordinate. In fact our argument remains valid whenever the Weyl anomaly of the dual CFT vanishes. The result has been proved in generality for even dimensional AdS by Fefferman and Graham for any choice of boundary metric. Since the Weyl anomaly for any CFT in odd number of dimensions vanish, this is a special case of our result. We will use our power series ansatz for the metric in Fefferman-Graham coordinates to show that the boundary stress tensor expectation value uniquely fixes all the coefficients in the power series thus specifying the solution uniquely. Given the CFT argument for the consistency of the power series ansatz we will be able to establish that the metric is uniquely determined locally by the stress tensor.  

It is clear, however, that any arbitrary traceless and conserved stress tensor will not correspond to a CFT state. For $AdS_5$ asymptotics we can say something more about gravity solutions with such boundary stress tensors. Even in these cases, we will prove rigorously that the power series solution with no $log$ terms in the radial coordinate exists when the boundary metric is flat. However in such gravity solutions either of two distinct pathologies can occur. For stress tensors with pathology of the first kind the reverse question of finding the corresponding gravity solution will be ill-posed. For such stress tensors, the formal power series solution of the metric in Fefferman-Graham coordinates will exist but this power series will have zero radius of convergence in the radial coordinate. These pathological stress tensors will be of the ``asymptotic boundary condition destroying'', or, in short, of ``abcd'' type. The other distinct set of pathological stress tensors will produce naked singularities in the bulk.

We will argue that ``abcd'' type of stress tensors can be avoided by doing a perturbation around a stationary late-time solution. We will further specialise to solutions with no net asymptotic angular momentum and these solutions at late times will always settle down to a static black brane. \footnote{This late time equilibriation, is of course expected only if the boundary stretches indefinitely in time, i.e. if the solution is free of ``abcd'' type of pathology. One may see this explicitly by studying an example, in which the boundary stress tensor is that of two fluids eternally moving past each other at different but constant velocities and temperatures without equilibriating. Our results imply that a solution with AdS asymptotics will exist even for such a boundary stress tensor. One of the authors (AM) is investigating this solution to check if it indeed has ``abcd'' type of pathology.} Multi blackbrane static solutions will not occur if there are no p-form gauge fields as is the case in pure gravity. We will set up a perturbation expansion in the Fefferman-Graham coordinates and show that all hydrodynamic stress tensors preserve the asymptotic AdS boundary condition. This result, we will argue, should also have some measure of validity for solutions carrying net angular momentum.

The perturbation expansion will be similar in spirit to that described in \cite{0712.2456,0803.2526}, but we will use Fefferman-Graham coordinates instead of Eddington-Finkelstein coordinates. A single black brane preserves the SO(3) rotation symmetries and the $\mathcal{R}^{3,1}$ translation symmetries of the full SO(4,2) isometries of $AdS_5$. Among the isometries which are broken only two can at most commute with each other and there is a four parameter family of choice of these two isometries. Since they parametrise the mutually commuting set of broken symmetries of the vacuum, which is the static black brane, we will call these ``maximally commuting Goldstone parameters''. We will choose them to be the scale transformation with one scaling parameter and an arbitrary boost parametrised by the three spatial components of a velocity. We can use them to generate a four parameter family of so-called boosted black brane solutions. This choice is natural because the boundary stress tensor of these boosted black brane solutions will be that of a homogenous perfect conformal fluid parameterised by its velocity and temperature. The velocity of the fluid will be the same velocity which parametrises the boost and the temperature will be the parameter of the scale transformation if the unboosted black brane had temperature unity (in units where the radius of AdS is set to unity). Now we will make the velocity and temperature arbitrary functions of the field theory coordinates (i.e all coordinates except the radial one) and find a correction to the metric which is first order in derivatives of the field theory coordinates. The boundary stress tensor is also corrected as a result and Einstein's equation implies it is conserved and traceless. This perturbation being an order by order derivative expansion should be thought of as the holographic dual of the usual low energy expansion (E/T) in an effective field theory, T being the temperature. This is therefore a derivative expansion.

The derivative expansion in the Fefferman-Graham has some advantages over the same expansion in Eddington-Finkelstein coordinates \cite{0712.2456,0803.2526}. In the Fefferman Graham coordinate system we can naturally view Einstein's equation as evolution of boundary metric in the radial direction. We will call those components of Einstein's equation which contain no more than one derivative of the radial coordinate as constraint equations. The first advantage is that the constraint equations become trivial except for the conservation and tracelessness of the boundary stress tensor if the dissipative (i.e the non-equilibrium) part of the boundary stress tensor $t_{(dis)\mu\nu}$ is chosen to satisfy $u^{\mu}t_{(dis)\mu\nu} =0$. The latter is called the Landau gauge condition and may be imposed without any loss of generality as by suitable redefinitions of the four velocity and temperature we can always make the stress tensor satisfy this property. \footnote{The Landau gauge is simply a convenient set of definitions of the velocity and temperature variables of the fluid and has nothing to do with gauge fixing of Einstein's equations. The physical meaning of these deifinitions is that $u^{\mu}$ is the local four-velocity of energy transport.} The second advantage over the perturbation in Eddington-Finkelstein coordinates is that here the whole procedure will be Lorentz-covariant, whereas in the Eddington-Finkelstein coordinates we had to decompose all terms into tensors, vectors and scalars of SO(3). The third advantage is that we can construct the metric for an arbitrary conformal hydrodynamic stress tensor.  We can also read off the stress tensor from our metric rather easily. Given this simplification of the constraints, in particular, one can think of the Fefferman-Graham coordinate system as the ``Coulomb gauge'' in the context of finding out metrics corresponding to arbitrary hydrodynamic stress tensors. 

However, as we already know from the results of \cite{0712.2456,0803.2526}, the solution corresponding to a generic hydrodynamic stress tensor will contain a naked singularity. In the Fefferman-Graham coordinates, however, we will find that the solution always has a singularity at the location of the unperturbed horizon. To see if the singularity is just a coordinate singularity or a real one we will translate our solution to Eddington-Finkelstein coordinates, because in the Eddington-Finkelstein coordinate system a real singularity will be manifest in terms of an actual blowup of the metric. To do this we will solve the equations of transformation exactly to each order in the derivative expansion. We will show that whether the singularity in the metric in Fefferman-Graham coordinates is real or fake, the translation to Eddington-Finkelstein coordinates can be achieved at every order. The metric in the Eddington-Finkelstein coordinates will make the singularity manifest and also easily reveal for which choice of the coefficients in the stress tensor would the solution be free of naked singularities. At every order in the derivative expansion, there will be a unique choice of coefficients of the terms in the stress tensor for which the solution will be free of naked singularities. 

Though we will establish the general results stated above, we will give explicit computations only upto first order in derivatives. In particular we will find the solution (exact upto first order in derivatives) in Fefferman-Graham coordinates for a conformal hydrodynamic stress tensor with arbitrary $\eta/s$. We will be able to find the solution for an arbitrary velocity configuration of the boundary fluid. A special case of our result will be the solution corresponding to the Bjorken flow found by Janik \cite{0703243,0512162}. With our method we will be able to find the solutions for arbitrary slowly varying velocity configurations at each order in the derivative expansion. It should also be kept in mind that the pathologies pointed out in \cite{0703243,0512162,0807.3797}, associated with the methods of finding solutions in Fefferman-Graham coordinates in \cite{0703243,0512162,0807.3797}, do not occur in our case because we never take a late time scaling limit in which we are zooming closer to the horizon, where in fact the metric always develops a coordinate singularity. In fact our method is as good and of equal reach as the derivative expansion in Eddington-Finkelstein coordinates. It has several comparative advantages which have been pointed out earlier, the comparative disadvantage being a slightly more elaborate regularity analysis. However if we go beyond the hydrodynamic sector to describe multi black brane solutions (if they exist), the Fefferman-Graham coordinate system (being tied to the AdS asymptotics) can always be employed efficiently, whereas it is not clear if the Eddington-Finkelstein coordinates will be equally useful.

The organisation of the paper is as follows. In section 2, we establish that the boundary stress tensor uniquely specifies a solution of pure classical gravity with AdS asymptotics when the boundary metric is flat. In section 3, we confirm our claims about the metric in Fefferman-Graham coordinates by translating a known solution in Eddington-Finkelstein coordinates which is exact upto first order in derivatives and free of naked singularities (we will call this solution as the hydrodynamic solution and has been found in \cite{0712.2456}). In section 4, we will set up and elucidate the derivative expansion in the Fefferman-Graham coordinates and establish that all hydrodynamic stress tensors preserve asymptotic AdS boundary condition. In section 5, we will do the regularity analysis of our solutions. Finally we will end with some discussion on the field-theoretic implications of our results.  

\section{How the boundary stress tensor fixes the solution}
In this section we will restrict our attention mainly to a five dimensional asymptotically AdS space with flat boundary metric, though we will indicate in the end that our results may be sufficiently generalised. We will soon explain what is meant by the boundary metric for asymptotically AdS spaces.

The Einstein-Hilbert action on 5-dim manifold $M$, with an appropriate counterterm to have a well defined variational principle with Dirichlet boundary condition is
\begin{equation}
S = \frac{1}{16\pi G_N}[-\int_{M} d^5x\sqrt{G}(R + \frac{12}{l^2})-\int_{\partial
    M}d^4x\sqrt{\gamma}2K]
\end{equation} 
where $K$ is the extrinsic curvature and $\gamma$ is the induced metric on the boundary. We are using the convention of \cite{0002230} in which the cosmological constant $\Lambda$ of $AdS_{d+1}$ is normalized to be $-\frac{d(d-1)}{2l^2}$, hence for $AdS_{5}$ we have $\Lambda= -\frac{6}{l^2}$.

We want to solve Einstein's equation 
\begin{equation}
R_{MN}- \frac{1}{2}RG_{MN}=\frac{6}{l^2}G_{MN}
\end{equation}
subject to the condition that the solution is asymptotically AdS with a given conformal structure at the boundary. Fefferman and Graham have shown that for such solutions we can use a specific coordinate system called the Fefferman-Graham coordinate system near the boundary. In this coordinate system, the metric takes the following form:
\begin{equation}\label{FG}
ds^2= G_{MN}dx^{M}dx^{N}=\frac{l^2}{\rho^2}[d\rho^2+g_{\mu\nu}(\rho,z)dz^{\mu}dz^{\nu}]
\end{equation}
In the expression above the indices (M,N) run over all AdS coordinates and the indices ($\mu,\nu$) run over the four field theory coordinates. The boundary metric $g_{(0)\mu\nu}$ is defined as 
\begin{equation}
g_{(0)\mu\nu}(z)= \lim_{\rho\rightarrow 0}g_{\mu\nu}(z,\rho)
\end{equation}
Let this boundary metric have a conformal structure. Then it can be shown that any conformal transformation of the boundary coordinates (z) can be lifted to a bulk diffeomorphism of the Fefferman-Graham coordinates which preserves the form of the metric (\ref{FG}) \cite{Brown:1986nw,penrose}. Under this bulk diffeomorphism, the boundary metric undergoes the same conformal transformation. The simplest case for instance will be a scale transformation, $z\rightarrow\lambda z$, of the boundary coordinates for which the corresponding bulk diffeomorphism will be $\rho\rightarrow\lambda\rho$ (note that in the case of the bulk diffeomorphism, the field theory coordinates z do not transform at all so that the boundary metric $g_{(0)\mu\nu}$ scales like $g_{(0)\mu\nu}(z)\rightarrow\lambda^{-2}g_{(0)\mu\nu}(z)$).

In the Fefferman-Graham coordinate system the various components of Einstein's equation reads as \cite{0002230}: \footnote{The (minor) difference with the system of equations given in this reference will be that we will use the original Fefferman-Graham radial coordinate $\rho$, whereas there the radial coordinate is chosen to be the squareroot of ours. Also, the reference uses a definition of the Riemann tensor such that the scalar curvature of AdS comes out to be positive.}
\begin{align}\label{eom}
\frac{1}{2}g^{\prime\prime }-\frac{3}{2\rho}g^{\prime}-\frac{1}{2}g^{\prime}g^{-1}g^{\prime}+\frac{1}{4}Tr(g^{-1}g^{\prime})g^{\prime}-Ric(g)-\frac{1}{2\rho}Tr(g^{-1}g^{\prime})g=0\\\nonumber
\nabla_{\mu}Tr(g^{-1}g^{\prime})-\nabla^{\nu}g^{\prime}_{\mu\nu}=0\\\nonumber
Tr[g^{-1}g^{\prime\prime}]-\frac{1}{\rho}Tr[g^{-1}g^{\prime}]-\frac{1}{2}Tr[g^{-1}g^{\prime}g^{-1}g^{\prime}]= 0
\end{align}
Here ``$(^\prime)$'' denotes a derivative with respect to $\rho$ and $\nabla_{\mu}$ is the covariant derivative constructed from the metric $g_{\mu\nu}$. Also in the above equations we have set our units such that l, the radius of AdS is set to unity.

When the boundary metric is flat, we will argue that we can expand $g_{\mu\nu}(z,\rho)$ in a simple integer power Taylor series of $\rho$ with coefficients which are functions of z. Since we have chosen the boundary metric to be flat, the leading term has to be $\eta_{\mu\nu}$. Our power series ansatz will be:
\begin{equation}\label{ans}
g_{\mu\nu}(z,\rho) =\eta_{\mu\nu} + \Sigma_{n=2}^{\infty}g_{(2n)\mu\nu}(z)\rho^{2n}
\end{equation}
We have written down only even powers of $\rho$ in the above expansion because it follows from a result due to Fefferman and Graham \cite{Fefferman} that the power series (\ref{ans}) should be an even function of $\rho$. \footnote{The existence of power series solution has been proved by Fefferman and Graham for all even dimensional asymptotic AdS solutions and in case of odd dimensional asymptotic $AdS$ solutions they also argued that if the solution is a power series it should be even. The Fefferman Graham coordinates are however unique only upto diffeomorphisms which are the lifts of the boundary conformal transformations into the bulk. Although, it is not obvious, it can also be shown \cite{Fefferman} that the evenness of the series (\ref{ans}) is independent of the choice of any particular Fefferman-Graham coordinate system.} The only even term which is absent is $g_{(2)\mu\nu}(z)$ which follows as an easy consequence of the equations of motion (\ref{eom}).

It is not obvious that this power series ansatz will indeed provide us a solution, so we will give an intuitive argument why this works. By AdS/CFT correspondence any solution of the bulk equations of motion would give us a state in the CFT, so the coefficients of the Taylor series expansion in (\ref{ans}) should be functions of the expectation values of the local operators in the dual CFT state. We will explicitly see below that all these coefficients are just functions of the expectation value of the stress tensor in the CFT state. It is possible to see the effect of space-time independent scale transformation on the CFT operators from $g_{\mu\nu}(z,\rho)$. To do this we have to lift the scale transformation to a bulk diffeomorphism so that the form of the metric (\ref{FG}) remains the same and the boundary metric also remains flat. This lift, as mentioned before, is achieved by $\rho\rightarrow\lambda\rho$. In the most general case it has been shown \cite{0209067} that the form of the ansatz (\ref{ans}) should be modified by terms like $\rho^{n}(log(\rho))^{m}$ with non-negative n and m. To illustrate our argument we will consider just two such possible terms:
\begin{equation}
g_{(n)}(z)\rho^{n}+h_{(n)}(z)\rho^{n}log(\rho) \nonumber
\end{equation}
Under the bulk scaling transformation {$\rho\rightarrow\lambda\rho$}, 
\begin{equation}\label{arg}
g_{(n)}(z)\rightarrow\lambda^{n-2}g_{(n)}(z)-log(\lambda)\lambda^{n-2}h_{(n)}(z) 
\end{equation}
We find the above transformation by checking the new coefficient of $\rho^n$ in $g_{\mu\nu}$ after the scale transformation. In a CFT any local operator simply scales like a power of $\lambda$, the power being given by the conformal dimension of the operator. A $log(\lambda)$ term is present only when the Weyl anomaly doesn't vanish. In flat space the Weyl anomaly vanishes and since we have chosen the boundary metric to be flat the $log$ term in (\ref{arg}) should not be present as $g_{(n)\mu\nu}$ is a function of the expectation values of local operators. The absence of the $log(\lambda)$ term in a scale transformation applies not only to primary operators but also to their descendents. So we can argue that terms like $\rho^{n}(log(\rho))^{m}$ should be absent and $g_{\mu\nu}$ should be given by a simple power series of $\rho$.

However, our argument, of course, breaks down if the boundary stress tensor does not correspond to any CFT state. In Appendix A, we have given the general proof of the existence of the power series solution for $AdS_5$ asymptotics, so that even for such cases we can state that the solution, is indeed, a power series. In fact we will explicitly see, that for all hydrodynamic stress tensors, whether they do or do not correspond to CFT states, the solutions are always power series. 

Now we will substitute our ansatz (\ref{ans}) in the equations of motion (\ref{eom}) and solve them order by order in powers of $\rho$. It is known from earlier work of Skenderis et.al. \cite{0002230} that the first term $g_{(4)\mu\nu}(z)$ is just the expectation value of the stress tensor. Briefly this is how it comes about to be so. Upto this order the first equation (the tensor equation) identically vanishes while the second and third equation of motion give:
\begin{eqnarray}\label{cons}
Tr(g_{(4)})=0\\\nonumber
\partial^{\mu}g_{(4)\mu\nu}=0 
\end{eqnarray}
Since the equations of motion by themselves cannot specify $g_{(4)}$ we need a data from the CFT to specify it subject to the above constraints. Most naturally $g_{(4)}$ is the traceless conserved stress tensor of the CFT. However we can also explicitly check this. An explicit calculation shows that $g_{(4)}$ is indeed the Balasubramanian-Kraus stress tensor \cite{9902121} which could be defined for any asymptotically AdS space. Hence we may write:
\begin{equation}
g_{(4)\mu\nu}=t_{\mu\nu} 
\end{equation}

With our ansatz (\ref{ans}) it turns out that all the other coefficients $g_{(2n)}$ ($n>2$) are fixed uniquely by the equations of motion in terms of $g_{(4)}$ and its derivatives (or in other words the stress tensor and its derivatives). We observe that the first and the third of the equations of motion (\ref{eom}) (i.e. the tensor and the scalar equations) are sufficient to solve for $g_{(n)}$. All the higher powers of the second of the equations of motion (\ref{eom}) (i.e the vector equation) identically vanishes on imposing the constraints (\ref{cons}) i.e. by imposing the tracelessness and the conservation of the stress tensor. It is not difficult to argue that this should be the case because it can be shown \cite{0002230} that the second (i.e the vector) equation of motion simply implies the conservation of the Brown-York stress tensor (which when regulated becomes the Balasubramanian-Kraus stress tensor) for an arbitrary constant $\rho$ hypersurface. Now the conservation of the Brown-York stress tensor at a given hypersurface is not independent of the same requirement for another hypersurface, because in the ADM-like formulation of the Einstein's equations if we satisfy our constraints at a given hypersurface in which our initial conditions are given the evolution (here in the radial coordinate $\rho$) automatically satisfies the constraints. The conservation of the Brown-York stress tensor at the boundary is already forced at leading order in $\rho$ of the vector equation of motion through (\ref{cons}). Hence we should expect that the vector equation should not impose any new constraints on the stress tensor given that the tensor and scalar equations specify all the coefficients uniquely and this is exactly what is borne out. In our proof in Appendix A, we show how the tensor, vector and scalar equations of motion turn out to be consistent with each other when we employ the power series ansatz. 

Below we give the a few of the the coefficients $g_{(n)\mu\nu}$
\begin{equation}
g_{(6)\mu\nu}=-\frac{1}{12}\Box t_{\mu\nu}\\\nonumber
\end{equation}
\begin{equation}
g_{(8)\mu\nu}=\frac{1}{2}t_{\mu}^{\phantom{\mu}\rho}t_{\rho\nu}-\frac{1}{24}\eta_{\mu\nu}(t^{\alpha\beta}t_{\alpha\beta})+\frac{1}{384}\Box^2t_{\mu\nu}\\\nonumber
\end{equation}
\begin{equation}
\begin{split}
g_{(10)\mu\nu}=&-\frac{1}{24}(t_{\mu}^{\phantom{\mu}\alpha}\Box
t_{\alpha\nu}+t_{\nu}^{\phantom{\nu}\alpha}\Box
t_{\alpha\mu})\\\nonumber
&+\frac{1}{180}\eta_{\mu\nu}t^{\alpha\beta}\Box
t_{\alpha\beta}+\frac{1}{360}t^{\alpha\beta}\partial_{\mu}\partial_{\nu}t_{\alpha\beta}\\\nonumber
&-\frac{1}{120}t^{\alpha\beta}(\partial_{\mu}\partial_{\alpha}t_{\beta\nu}+\partial_{\nu}\partial_{\alpha}t_{\beta\mu})\\\nonumber
&+\frac{1}{60}t^{\alpha\beta}\partial_{\alpha}\partial_{\beta}t_{\mu\nu}-\frac{1}{180}\partial_{\mu}t^{\alpha\beta}\partial_{\nu}t_{\alpha\beta}\\\nonumber
&+\frac{1}{720}\eta_{\mu\nu}\partial_{\alpha}t^{\beta\gamma}\partial^{\alpha}t_{\beta\gamma}\\&+\frac{1}{120}(\partial_{\mu}t^{\alpha\beta}\partial_{\alpha}t_{\beta\nu}+\partial_{\nu}t^{\alpha\beta}\partial_{\alpha}t_{\beta\mu})\\\nonumber
&-\frac{1}{60}\partial_{\alpha}t^{\beta}_{\mu}\partial_{\beta}t_{\nu}^{\alpha}-\frac{1}{23040}\Box^3t_{\mu\nu}\\\nonumber
\end{split}
\end{equation}
\begin{equation}\label{coeff}
g_{(12)\mu\nu}= \frac{1}{6}t_{\mu}^{\phantom{\mu}\alpha}t_{\alpha}^{\phantom{\alpha}\beta}t_{\beta\nu} - \frac{1}{72}t_{\mu\nu}(t^{\alpha\beta}t_{\alpha\beta}) + ........
\end{equation}
Here, as before in (\ref{eom}) the boundary indices are raised and lowered by $\eta_{\mu\nu}$ and $\Box$ is the Laplacian in flat space. Let us observe and explain certain simple features of the results above. The first observation is that every term in the RHS of the above equations contain only even number of derivatives. This is so because the terms containing derivatives originate only from Ric(g) in the first of the equations of (\ref{eom}). The second observation is that the terms independent of the derivatives appear only for $g_{(4n)}$. This is so because if we omit Ric(g) in the first of the equations of (\ref{eom}), then the solution is a power series in $\rho^{4n}$ as the first non-trivial term in the series is $g_{(4)}$. So for a solution where the stress tensor is uniform (like in the case of a static black brane solution), g has an expansion containing only $\rho^{4n}$ terms.

With our argument that the ansatz (\ref{ans}) should give us a consistent solution, it is obvious that the stress tensor, which appears as $g_{(4)}$ in g uniquely specifies the solution because all the higher coefficients are fixed uniquely in terms of $g_{(4)}$ with no new constraints like (\ref{cons}) appearing for $g_{(4)}$. This completes the argument that when the boundary metric is flat we should have a solution uniquely specified locally by the stress tensor alone. This statement readily generalizes to other dimensions in the case of a flat boundary metric and most likely also generalizes when the boundary metric is not flat. The general validity could be argued for on the basis of the equations of motion (\ref{eom}) which are second order (specifically in derivatives of $\rho$). Intuitively the boundary metric and the stress tensor specifies all the initial data we need for a unique solution, however a concrete demonstration of this would probably require methods beyond what we have employed here.

Our power series ansatz (\ref{ans}) should work even if the Einstein-Hilbert action with negative cosmological constant receives higher derivative corrections provided the boundary stress tensor corresponds to a state in the dual theory. Our argument as to why it works is independent of the equation of motion and likewise also independent of say, the value of t'hooft coupling of the dual theory. We have just used the fact that a conformal transformation in the boundary should have an appropriate lift to a bulk diffeomorphism consistent with the transformation of CFT operators. The transformation of the CFT operators under conformal transformations, as well, is independent of the value of the coupling. In fact one can readily check that exact static black hole solutions of Gauss-Bonnet gravity which are asymptotically AdS (given in \cite{Cai}) have power series expansion when written in Fefferman-Graham coordinates.

The argument we have given above, however, cannot be reversed to argue that a solution with asymptotic $AdS_5$ boundary conditions exists for any arbitrary stress tensor. The reason that we can't reverse the argument is that the series (\ref{ans}) for $g_{\mu\nu}$ exists only formally. The coefficients $g_{(n)}$ may not be well behaved at large n, for an arbitrary stress tensor. We will give a simple example to show what can go wrong. For a specific choice of stress tensor, we may find that $g_{(n)\mu\nu} = f(n)s_{\mu\nu}$ plus other terms. Here $s_{\mu\nu}$ is a specific term in the stress tensor. If, for instance, the series $\Sigma_{n} f(n)\rho^{n}$ has zero radius of convergence, $g_{\mu\nu}$ will not be a meaningful series of $\rho$ as it will also have zero radius of convergence in $\rho$. Such boundary stress tensors, for which $g_{\mu\nu}$  has zero radius of convergence in $\rho$, could be appropriately called, ``asymptotic boundary condition destroying'' stress tensor or in short ``abcd'' stress tensor. We will have more to say about such stress tensors in section 4.\footnote{Interestingly, Fefferman and Graham have shown in \cite{Fefferman} that for even dimensional asymptotic AdS solutions, $g_{\mu\nu}$ always has a finite radius of convergence in $\rho$. However their argument does not readily generalize to the odd dimensional case.}
 
\section{Mutual translation between Eddington-Finkelstein and Fefferman-Graham coordinates}
In the previous section, we have seen that, the Fefferman-Graham coordinate system is good for finding a solution to Einstein's equation with a negative cosmological constant when the corresponding boundary stress tensor is specified. However the solutions are usually found in other coordinate systems. For instance, the static black brane solution is usually described in the Schwarzchild-like coordinate system and the hydrodynamic metric of \cite{0712.2456} has been found in the Eddington-Finkelstein coordinate system. It would be useful to see how we can rewrite these solutions in the Fefferman-Graham coordinate system asymptotically. We will demonstrate a novel technique towards this end for the boosted black brane and the hydrodynamic metrics. In both cases we will see that we can achieve a mutual translation between Eddington-Finkelstein coordinate system and Fefferman-Graham coordinate system by using a power series ansatz similar to (\ref{ans}) and we can solve this ansatz algebraically order by order. We expect this method to work for all solutions in which the boundary metric is flat, or more generally when the Weyl anomaly vanishes.

The general procedure is as follows. In the Eddington-Finkelstein coordinates ($x^{\mu}, r$)  the metric takes the form:
\begin{equation}\label{EF}
ds^2 = -2u_{\mu}(x)dx^{\mu}dr+ G_{\mu\nu}(x,r)dx^{\mu}dx^{\nu}
\end{equation}
Here we are using ingoing Eddington-Finkelstein coordinate system, so that $u^{\mu}$ is a four-velocity (hence $u_{\mu}u_{\nu}\eta^{\mu\nu}=-1$) such that it is directed forward in time. We will express the general structure of coordinate transformation from the Eddington-Finkelstein coordinates ($x^{\mu}, r$) to Fefferman-Graham coordinates ($z^{\mu}, \rho$) as below: 
\begin{eqnarray}\label{defn}
d{\rho}=p_{\mu}(r,x)dx^{\mu}+q(r,x)dr\\
dz^{\mu}=m^{\mu}_{\phantom{\mu}\nu}(r,x)dx^{\nu}+n^{\mu}(r,x)dr
\end{eqnarray}
We substitute the above in the Fefferman-Graham form of the metric (\ref{FG}) to get:
\begin{equation}
\begin{split}
ds^2=\frac{1}{{\rho}^2}&[(p_{\mu}p_{\nu}+g_{\eta\xi}(\rho,z)m^{\eta}_{\phantom{\eta}\mu}m^{\xi}_{\phantom{\xi}\nu})dx^{\mu}dx^{\nu}+2(p_{\mu}q+g_{\xi\sigma}(\rho,z)m^{\xi}_{\phantom{\xi}\mu}n^{\sigma})dx^{\mu}dr\\&+(q^2+g_{\mu\nu}(\rho,z)n^{\mu}n^{\nu})dr^2]
\end{split}
\end{equation}
Comparing the above with the Eddington-Finkelstein form of the metric (\ref{EF}), we get the following set of equations:
\begin{eqnarray}\label{trans}
(q(x,r))^2 + g_{\mu\nu}(\rho, z)n^{\mu}(x,r)n^{\nu}(x,r) = 0\\\nonumber
2p_{\mu}(x,r)q(x,r) + g_{\alpha\beta}(\rho, z) (m^{\alpha}_{\phantom{\alpha}\mu}(x,r) n^{\beta}(x,r) +  m^{\beta}_{\phantom{\beta}\mu}(x,r) n^{\alpha}(x,r)) \\\nonumber= -2u_{\mu}(x)(\rho(x,r))^2\\\nonumber
p_{\mu}(x,r)p_{\nu}(x,r)+g_{\alpha\beta}(\rho,z)m^{\alpha}_{\phantom{\alpha}\mu}(x,r)m^{\beta}_{\phantom{\beta}\nu}(x,r) = G_{\mu\nu}(x,r)(\rho(x,r))^2 
\end{eqnarray}
So we have a scalar, a vector and a tensor equation and three unknowns to solve for. The unknowns are a scalar $\rho(x,r)$, a vector $z^{\mu}(x,r)$ and the tensor $g_{\mu\nu}(z,\rho)$ which appear in the Fefferman-Graham metric (\ref{FG}). It is clear from the definitions (\ref{defn}) of q, etc. that they are just various partial derivates of ($\rho, z$), for instance $q = \partial_{r}\rho$, etc. We will make the following general ansatz to solve the above equations. The ansatz for $\rho$ and $z^{\mu}$ will be that they will be an integer power series of the inverse of the Eddington-Finkelstein radial coordinate r.
\begin{eqnarray}\label{ans2}
\rho= \frac{1}{r}+\frac{{\rho}_{2}(x)}{r^2}+\frac{{\rho}_{3}(x)}{r^3}+.........\\\nonumber
z^{\mu}=x^{\mu}+\frac{z_1^{\mu}(x)}{r}+\frac{z_2^{\mu}(x)}{r^2}+.....
\end{eqnarray}
To solve the equations of transformation (\ref{trans}), the above should be supplemented with the ansatz (\ref{ans}) for the $g_{\mu\nu}(z,\rho)$ in the Fefferman Graham metric. The expressions for the partial derivatives like q, etc. then turn out to be as below:
\begin{eqnarray}
q=\partial_r{\rho}=-\frac{1}{r^2}-\frac{2{\rho}_2}{r^3}-\frac{3{\rho}_3}{r^4}-....\\\nonumber
p_{\mu}=\partial_{\mu}{\rho}=\frac{\partial_{\mu}{\rho}_2}{r^2}+\frac{\partial_{\mu}{\rho}_3}{r^3}+.....\\\nonumber
n^{\mu}=\partial_rz^{\mu}=-\frac{z_1^{\mu}}{r^2}-\frac{2z_2^{\mu}}{r^3}-....\\\nonumber
m^{\mu}_{\phantom{\mu}\nu}=\partial_{\nu}z^{\mu}=\delta^{\mu}_{\nu}+\frac{\partial_{\nu}z_1^{\mu}}{r}+\frac{\partial_{\nu}z_2^{\mu}}{r^2}+....
\end{eqnarray}
One thing to be kept in mind is that when we substitute our ansatz (\ref{ans2}) to solve the equations of transformation (\ref{trans}), $g_{\mu\nu}(\rho ,z)$ should be re-expressed as functions of (x,r). Below, we just give the first three terms which appear after it is rewritten as functions of (x,r).
\begin{equation}
g_{\mu\nu} = \eta_{\mu\nu} + \frac{t_{\mu\nu}(x)}{r^4} + \frac{(4\rho_{2}t_{\mu\nu}+(z_{1}.\partial)t_{\mu\nu})(x)}{r^5} +....
\end{equation}
We now consider a boosted black brane metric in Eddington-Finkelstein coordinate
\begin{equation}\label{bbm}
ds^2=-2u_{\mu}dx^{\mu}dr-r^2f(br)u_{\mu}u_{\nu}dx^{\mu}dx^{\nu}+r^2P_{\mu\nu}dx^{\mu}dx^{\nu}
\end{equation}
where
\begin{eqnarray}
f(r)=1-\frac{1}{r^4}\\
u^0=\frac{1}{\sqrt{1-\beta_i^2}}\\
u^i=\frac{\beta_i}{\sqrt{1-\beta^2_i}}
\end{eqnarray}
and the temperature is $T=\frac{1}{\pi b}$ and the three-velocity $\beta_i$ are all constants, and
\begin{equation}
P_{\mu\nu}=u_{\mu}u_{\nu}+\eta_{\mu\nu}
\end{equation}
is the projector onto the spatial hypersurface orthogonal to the four velocity $u^{\mu}$. This metric can be obtained by applying a boost parameterised by the three-velocity $\mathbf{\beta}_{i}$ and a scaling by $b$ to the usual AdS black hole with unit temperature where the time coordinate $t$ is itself a Killing vector. In this case actually the exact transformation from Eddington-Finkelstein to Fefferman-Graham coordinate system can be exactly worked out easily and it is given by:
\begin{eqnarray}\label{bb}
\rho = \frac{\sqrt{2}b}{\sqrt{b^2 r^2 + \sqrt{b^4 r^4 -1}}}\\\nonumber
z^{\mu} = x^{\mu} + u^{\mu}bk(br),\\\nonumber
k(y) = \frac{1}{4}(log(\frac{y+1}{y-1}) - 2 arctan(y) + \pi) 
\end{eqnarray}
The solution for $g_{\mu\nu}$ in the Fefferman-Graham metric (\ref{FG}) for the boosted black brane is given by:
\begin{equation}\label{go}
g_{\mu\nu}(z,\rho) = (1+\frac{{\rho}^4}{4b^4})\eta_{\mu\nu}+\frac{4{\rho}^4}{4b^4+{\rho}^4}u_{\mu}u_{\nu}
\end{equation}
The boundary stress tensor could be easily read off by looking at the coefficient of $\rho^4$ after Taylor expanding the RHS of the above expression. The stress tensor turns out to be that of an ideal conformal fluid (like that of a gas of photons)
\begin{equation}\label{pf}
t_{0\mu\nu}={g_{(4)}}_{\mu\nu}=\frac{1}{4b^4}[4u_{\mu}u_{\nu}+\eta_{\mu\nu}]
\end{equation}
where the temperature is $T=\frac{1}{\pi b}$. The horizon in the Fefferman-Graham coordinates is at $\rho=\sqrt{2}b$ and at the horizon $g_{\mu\nu}$ given by (\ref{go}) is not invertible as $g_{\mu\nu}(\rho=\sqrt{2}b,z) = 2 P_{\mu\nu}$. So clearly the Fefferman-Graham coordinate system has a coordinate singularity at the horizon. Also it is easy to check from (\ref{bb}) that the change of coordinates also becomes singular at the horizon.

Now we turn to the hydrodynamic metric found in \cite{0712.2456} which is a solution to Einstein's equation upto first order in the derivative expansion and has a regular horizon. Here the ``maximally commuting Goldstone parameters'' of the boosted black brane solution, the velocities $\beta^{i}$ and the temperature $T$ are functions of the field theory coordinates (x). The $G_{\mu\nu}$ in the Eddington-Finkelstein form of the metric (\ref{EF}) is:
\begin{equation}
G_{\mu\nu} = r^2 P_{\mu\nu} + (-r^2 +\frac{1}{b^4 r^2})u_{\mu}u_{\nu} + 2r^2 bF(br)\sigma_{\mu\nu} -r((u.\partial)u_{\mu}u_{\nu}-\frac{2}{3}u_{\mu}u_{\nu}(\partial .u)) 
\end{equation}
with
\begin{equation}
F(x) = \frac{1}{4}(log(\frac{(x+1)^2(x^2+1)}{x^4}) - 2 arctan(x) + \pi)
\end{equation}
In this case we will solve the set of equation (\ref{trans}) by putting in our anstaz (\ref{ans2}). We solve order by order for each power $n$ in $r^{-n}$. At each order we have to solve algebraic equations and remarkably the equations can be consistently solved at each order. It is important to throw away all the terms which have two $x$-derivatives or more and solve the series for $\rho$ and $z^{\mu}$ given in (\ref{ans2}) and the series for $g_{\mu\nu}$ given in (\ref{ans}) only upto first derivative order. This is justified because the hydrodynamic metric above in Eddington-Finkelstein form is a solution to Einstein's equation only upto first order in $x$-derivatives and hence it can have a Fefferman-Graham expansion near the boundary only upto first derivative order. The results of the non-vanishing terms in the expansion for $\rho$ and $z^{\mu}$ in (\ref{ans2}) upto $r^{-9}$ order are given below:
\begin{eqnarray}\label{sol}
\rho_2 = \frac{1}{3}(\partial . u), {\rho}_5=\frac{1}{8b^4}, \rho_6 = \frac{13(\partial.u)}{120 b^4}, {\rho}_9=\frac{7}{128b^8}\\\nonumber
z_{1}^{\mu}=u^{\mu}, z_2^{\mu} = \frac{1}{3}u^{\mu}(\partial.u), z_{5}^{\mu}=\frac{u^{\mu}}{5b^4},\\\nonumber
z_6^{\mu} =  \frac{9u^{\mu}(\partial.u)+7(u.\partial)u^{\mu}}{60b^4}, z_{9}^{\mu}=\frac{u^{\mu}}{9b^8}
\end{eqnarray}
We can easily observe some patterns in the results above. Firstly the terms without any derivatives only appear as coefficients of $r^{-4n-1}$. These are precisely the terms that appear in the expansion for the case of the boosted black brane as given in (\ref{bb}). This is because the original black brane solution in Fefferman-Graham coordinates as we know from (\ref{go}) is a series with ``gaps'' of four (which means only the fourth next term is non-zero). So the solution of (\ref{trans}) should provide a series for $\rho$ and $z^{\mu}$ in gaps of four as well. Secondly, it also turns out that the terms which have first derivative pieces occur for $\rho_2 , \rho_6, z_2^\mu, z_6^\mu$, etc. again in gaps of four. We obtain the coefficients of the series for $g_{\mu\nu}$ given in (\ref{ans}) which was part of our ansatz. The second non-zero term in the series gives us the boudary stress tensor:
\begin{equation}\label{td}
t_{\mu\nu}=g_{(4)\mu\nu}= \frac{\eta_{\mu\nu} + 4u_{\mu}u_{\nu}}{4b^4}-\frac{1}{2 b^3}\sigma_{\mu\nu}
\end{equation}
where
\begin{equation}\label{dis}
\sigma_{\mu\nu}=P_{\mu}^{\phantom{\mu}\alpha}P_{\nu}^{\phantom{\nu}\beta}\partial_{(\alpha}u_{{\beta})}-\frac{1}{3}P_{\mu\nu}\partial_{\alpha}u^{\alpha}
\end{equation}
This is stress tensor for a relativistic conformal fluid satisfying Navier-Stokes' equation and with $\eta/s=1/4\pi$. The next non vanishing term in the series for $g_{\mu\nu}$ is:
\begin{equation}
g_{(8)\mu\nu} = -\frac{u_{\mu}u_{\nu}}{4b^8}-\frac{\sigma_{\mu\nu}}{8b^7}
\end{equation}
We can check that the expression for $g_{(8)}$ is given by the general results of the the previous section when we substitute the dissipative stress tensor (\ref{td}) in (\ref{coeff}).

In this section we have worked out the case for a specific ``hydrodynamic metric'' given in \cite{0712.2456}. This metric has no naked singularities and this corresponds to the choice of $\eta/s = 1/4\pi$ in the dissipative stress tensor (\ref{dis}). However we will see in section 5 that our ansatz (\ref{ans2}) for translation between the Eddington-Finkelstein and Fefferman-Graham coordinates will work even when the above is not the case, i.e the metric contains naked singularities. In what follows we will reverse the translation. That is, we will work out the Fefferman-Graham form of the metric exactly upto first order in derivatives first and then find out the Eddington-Finkelstein form of the metric also exactly upto first order in derivatives. We will see that the power series ansatz (\ref{ans2}) is consistent for any metric corresponding to an arbitrary hydrodynamic stress tensor.

\section{The derivative expansion in Fefferman-Graham coordinates}
We have already seen that the Fefferman-Graham form of the metric is the ideal one to use if we are asking given a boundary stress tensor what the corresponding solution of Einstein's equations of motion should be. The most general hydrodynamic stress tensor for a conformal fluid (in the Landau gauge) upto first order in derivatives is as below:
\begin{equation}\label{fost}
t_{\mu\nu}(z) = \frac{\eta_{\mu\nu} + 4u_{\mu}(z)u_{\nu}(z)}{4b(z)^4} - \frac{\gamma}{2 b(z)^3}\sigma_{\mu\nu}(z)
\end{equation}
with $\sigma_{\mu\nu}(z)$ given by (\ref{dis}), $b$ related to the temperature through $b=1/\pi T$ and $\gamma$ an arbitrary constant. However here, unlike in the case of the specific solution (without naked singularities) we considered in the previous section, $\eta/s =\gamma/4\pi$  and hence is arbitrary. We now ask what would be the corresponding solution for this arbitrary case. 

Before we get into this specific case, we will show that we can get some insights into the reverse question from some generally known facts and our previous results given in section 2. We have seen, briefly, at the end of section 2 that the reverse question is ill posed for an ``abcd'' (asymptotic boundary condition destroying) stress tensor, for which the formal power series (\ref{ans}) for $g_{\mu\nu}$ has zero radius of convergence in $\rho$. One must devise a strategy in which such stress tensors do not appear at all. To this end we may always exploit a general property of solutions of Einstein's equation that in the long run the solution always becomes stationary. For the moment let us further restrict to those solutions which have no (ADM) angular momentum or any other (ADM) conserved charges (like the R-charge). These will, in the long run, settle down to the known boosted black brane solution (\ref{bbm}). Static multi blackbrane like solutions do not appear if we turn off p-form gauge fields, so if more than one black brane are present they eventually will collapse to form a single black brane. A good strategy to recover all solutions will be to perturb around the late-time static black brane and build up all solutions in a systematic derivative expansion. Since any solution would eventually become static (or equilibriate) this strategy should always work at sufficiently late times.

Since the approach to equilibrium can be naturally described by hydrodynamics, one can intuitively expect that the late time behaviour of the solutions will correspond to a hydrodynamic description in terms of the boundary theory \textit{if the equilibrium can be described in terms of a perfect fluid}. The boundary stress tensor of a boosted black brane indeed corresponds to that of a perfect conformal fluid like that of photons in pure QED. Our expectation is indeed borne out by the fact that all solutions in the derivative expansion correspond to a traceless conserved hydrodynamic boundary stress tensor, but with arbitrary number of derivatives. We will see that in the derivative expansion at each order the solutions always have finite radius of convergence away from the boundary, so we can conclude that all hydrodynamic stress tensors are asymptotic boundary condition preserving.

The fact that all hydrodynamic stress tensors preserve the asymptotic AdS boundary condition should have a certain measure of validity even for solutions with net angular momentum. In fact in \cite{0708.1770}, it has been shown that a large class of rotating black holes in AdS can be described by perfect fluid hydrodynamics. However, we do not know how general the result is. The argument in the previous paragraph shows that for any solution if the hydrodynamic description holds for the stationary solution to which a given solution eventually equilibrates, it should hold for sufficiently late times as well. So certainly a large class of solutions even in the sector with net angular momentum which can be constructed by perturbing around certain stationary solutions will have a hydrodynamic description at least at late times. \footnote{As we have mentioned in a previous footnote, a non-trivial check of this strategy will be to construct a solution for a boundary stress tensor for which there is no late time equilibriation and see how it is connected to the ``abcd'' type of pathology.}

To build up a solution corresponding to an arbitrary hydrodynamic stress tensor, we will work in the Fefferman-Graham coordinate system as we have said before and we will construct the solution exactly order by order in the derivative expansion. To develop the derivative expansion we follow the same method which the authors of \cite{0712.2456} followed but now in the Fefferman-Graham coordinate system. In fact, based on the results of section 2, we will see that their method simplifies in these coordinates. We take the boosted black brane solution with $g_{\mu\nu}$ of the form of (\ref{go}), but now the ``maximally commuting Goldstone parameters'' ($u^{\mu}$, $b$) are arbitrary functions of z. We will call this the zeroth order metric $g_{0}$ which is no more a solution to Einstein's equation, so we need to correct this with $g_{1}$ which will now depend on the first derivatives of the ``maximally commuting Goldstone parameters'' ($u^{\mu}$, $b$). This correction $g_{1}$ can be found substituting $g=g_0 + g_1$ in our equations of motion (\ref{eom}) and retaining only terms which have no more than one derivative of z. 

The first of the equations of motion (\ref{eom}), i.e the tensor equation gives us a source free linear equation for $g_{1}$ which is second order in the derivatives of $\rho$ and has no z-derivatives. 
\begin{equation}\label{te}
\begin{split}
&\frac{1}{2}g_{1}^{''}-\frac{3}{2}\frac{g_{1}^{'}}{\rho} -\frac{1}{2}g_{1}^{'}g_{0}^{-1}g_{0}^{'}-\frac{1}{2}g_{0}^{'}g_{0}^{-1}g_{1}^{'}+\frac{1}{2}g_{0}^{'}g_{0}^{-1}g_{1}g_{0}^{-1}g_{0}^{'}\\&+\frac{1}{2}(Tr(g_{0}^{-1}g_{1}^{'})-Tr(g_{0}^{-1}g_{1}g_{0}^{-1}g_{0}^{'}))(\frac{g_{0}^{'}}{2}-\frac{g_{0}}{\rho})+\frac{1}{2}Tr(g_{0}^{-1}g_{0}^{'})(\frac{g_{1}^{'}}{2}-\frac{g_{1}}{\rho}) = 0
\end{split}
\end{equation}
At the first order in derivative expansion, the only term which can provide a source term is Ric(g) since it has no derivatives of $\rho$. However Ric(g) contains at least two derivatives of z, so at this order the source vanishes. 

At the first order the second of the equations of motion, which is a vector equation gives us the following:
\begin{equation}
\nabla_{0\mu}Tr(g_{0}^{-1}g_{0}^{\prime})-\nabla_{0}^{\nu}g^{'}_{0\mu\nu}=0
\end{equation}
where $\nabla_0$ is the covariant derivative constructed from $g_0$.The major simplification which occurs in the Fefferman-Graham coordinates is the general observation in section 2, that this gives us nothing but the conservation of the stress tensor. It may be checked that if we choose to solve this vector fluctuation equation order by order in powers of $\rho$, like we did in section 2, at the leading order we would get $\partial^{\mu}t_{o\mu\nu} = 0$, where $t_{0\mu\nu}$ is the perfect fluid stress tensor (\ref{pf}) and all the coefficients of the higher powers of $\rho$ will vanish identically once the leading order condition is imposed. This simplification will happen at every order in the derivative expansion, which means that if $t_{n-1}$ is the stress tensor upto n-1 th order in the derivative expansion, at the n-th order the second equation will simply imply the conservation of $t_{n-1}$. 

At the first order in the derivative expansion the third equation of motion vanishes identically. It is easy to see why this will happen. Again we go back to the general observations of section 2. If $t_{\mu\nu} = t_{0\mu\nu}+t_{1\mu\nu}$  with $t_{0\mu\nu}$ given by the perfect fluid stress tensor (\ref{pf}) and $t_{1\mu\nu}$ is the first order correction to the stress tensor satisfying the tracelessness and the Landau gauge $u^{\mu}t_{1\mu\nu}=0$ conditions, then the correction to the coefficients of the power series expansion $g_{(n)\mu\nu}$ (some of which are listed in (\ref{coeff})) is simply proportional to $t_{1\mu\nu}$. The first order derivatives of $t_{0\mu\nu}$ doesn't appear because, as we have observed the general expressions for $g_{(n)}$ must contain even number of derivatives of $t_{0\mu\nu}$. It follows that the correction to the zeroth order metric, $g_{1}$, is proportional to $t_{1}$. It also follows from the the tracelessness of $t_{1}$ and the Landau gauge condition that the third equation vanishes identically as all traces appearing in the equation vanish. We will soon see that, this simplifying feature also, remarkably generalises to all orders in the derivative expansion. 

In the Fefferman-Graham coordinates the first order correction to the metric $g_{1}$ is, therefore, proportional to the first order correction to the stress tensor which is proportional to $\sigma_{\mu\nu}$ and therefore $g_1$ takes the form of $\gamma^{'} b \sigma_{\mu\nu} f(\rho)$, where $\gamma^{'}$ is an arbitrary constant. Substituting this in the tensor equation (\ref{te}), we find that $f(\rho)$ satisfies the following differential equation:
\begin{equation}\label{eof}
f^{''} - f^{'}\frac{(12b^4 - \rho^4)(4b^4 + 3\rho^4)}{\rho(16b^8 - \rho^8)} + f \frac{128\rho^6 b^4}{(4b^4+\rho^4)(16b^8 - \rho^8)}=0 
\end{equation}
We already know that the solution is a power series in $\rho^4$, so we change our variable $\rho$ to $x = \rho^4$. The equation now reads
\begin{equation}
f^{''} - f^{'} \frac{8b^4}{16b^8 - x^2} + f \frac{8b^4}{(4b^4 + x)(16b^8 - x^2)}=0
\end{equation}
The solution of this differential equation which vanishes at the boundary (after resubstituting $x$ with $\rho^4$) \footnote{The other solution is $f_2 = 1 + \frac{\rho^4}{4b^4}$} is:
\begin{equation}\label{f}
(1+\frac{\rho^4}{4b^4})log(\frac{1-\frac{\rho^4}{4b^4}}{1+\frac{\rho^4}{4b^4}})
\end{equation}
The metric in Fefferman-Graham coordinates upto first order then is:
\begin{equation}
ds^2 = \frac{d\rho^2 + g_{\mu\nu}(\rho, z)dz^{\mu}dz^{\nu}}{\rho^2}\\\nonumber
\end{equation}
\begin{equation}\label{mfo}
g_{\mu\nu}(\rho, z) = (1+\frac{\rho^4}{4b^4})\eta_{\mu\nu} + \frac{4\rho^4}{4b^4 + \rho^4}u_{\mu}u_{\nu} + \gamma^{'} b \sigma_{\mu\nu}(1+\frac{\rho^4}{4b^4})log(\frac{1-\frac{\rho^4}{4b^4}}{1+\frac{\rho^4}{4b^4}})
\end{equation}
To read off the stress tensor upto first order, we simply need the $\rho^{4}$ term in the Taylor expansion of $g_{\mu\nu}$. We get:
\begin{equation}
t_{\mu\nu} = \frac{\eta_{\mu\nu} + 4u_{\mu}u_{\nu}}{4b^4} - \frac{\gamma^{'}}{2 b^3}\sigma_{\mu\nu}
\end{equation}
Comparing with (\ref{fost}) we get that we must set $\gamma^{'} = \gamma$ in the first order metric (\ref{mfo}) to get the desired solution corresponding to the boundary stress tensor.

One very interesting feature of our solution at the first order can be found out by putting $\gamma^{'} = \gamma =0$. This implies that our zeroth order solution itself, now with velocities and temperatures satisfying the relativistic Euler equation, is an exact solution of Einstein's equations upto first order. Such is never the case in Eddington-Finkelstein coordinate system where as we will see we need to correct the zeroth order solution even for a dissipation-less stress tensor so that the solution is exact upto first order. We do not understand any deep reason for this feature of our solution.

Now we can proceed to examine the higher orders in the derivative expansion. Though we will postpone explicit solutions beyond the first order for a future publication, here we will show that it is trivial to satisfy the vector and scalar constraints at each order in perturbation theory. The tensor equation takes the following form at each order in perturbation theory:
\begin{equation}\label{tef}
\begin{split}
&D_1g_{n\mu\nu}+D_2(g_{n\mu\rho}u^{\rho}u_{\nu}+g_{n\nu\rho}u^{\rho}u_{\mu})+D_3(g_{n\rho\sigma}\eta^{\rho\sigma})\eta_{\mu\nu}+D_4(g_{n\rho\sigma}\eta^{\rho\sigma})u_{\mu}u_{\nu}\\ &+D_5(g_{n\rho\sigma}u^{\rho}u^{\sigma})\eta_{\mu\nu}+D_6(g_{n\rho\sigma}u^{\rho}u^{\sigma})u_{\mu}u_{\nu}=s_{n\mu\nu}(z,\rho)
\end{split}
\end{equation}
where $D_1$, $D_2$, etc. are linear differential operators involving derivatives in the radial coordinate only and $s_{n\mu\nu}(z,\rho)$ is the source term which is a (nonlinear) function of the corrections to the metric upto n-1 th order in the derivative expansion. The left hand side of the above equation is in fact the same as in (\ref{te}) with $g_1$ replaced by the n-th order correction to the metric $g_n$, but now source terms are present on the right hand side. Also the differential operator $D_1$ is the same as the operator which acts on f in (\ref{eof}) at every order in the derivative expansion. We dropped the operators $D_2$, $D_3$, etc. at the first order, i.e. for $g_1$, because as we saw the general results of section 2 (equations in (\ref{coeff}) for instance) forced it to be proportional to be stress tensor and hence be traceless and vanish when contracted with the four velocity. However, from the second order in the derivative expansion onwards, the general results of section 2 do not imply this to be true for the correction to the metric and in fact the source terms which appear on the right hand side of the equation indeed do not have this property. All the other operators except $D_1$, however, involve no more than one derivative in the radial coordinate.

We have to choose a particular solution to the above equation. We can always choose the particular solution to be such that it vanishes at the boundary like $\rho^6$ so that it doesn't contribute to the stress tensor (as the coefficient of its $\rho^4$ term vanishes). One can explicitly check this, however, more efficiently we can prove it as follows. The source term for the n-th order correction clearly is determined by various terms of the stress tensor upto n-1 th order, so it follows from the general results of section 2 that the particular solution can be chosen to be independent of $t_{n\mu\nu}$, which is the n-th order correction to the stress tensor. In that case the $\rho^4$ term should be absent. For instance, based on the results like those in (\ref{coeff}), we can write down the Taylor series expansion in the radial coordinate for the particular solution for $g_2$ as below.
\begin{equation}
\begin{split}
&g_{2\mu\nu}=-\frac{\rho^6}{12}\Box t_{0\mu\nu}+
\rho^8[\frac{1}{2}t^{\phantom{\mu}\rho}_{1\mu}t_{1\rho\nu}-\frac{1}{24}\eta_{\mu\nu}(t_{1}^{\rho\sigma}t_{1\rho\sigma})] \\
&+\rho^{10}[-\frac{1}{24}(t_{0\mu}^{\phantom{\mu}\alpha}\Box
t_{0\alpha\nu}+t_{0\nu}^{\phantom{\nu}\alpha}\Box
t_{0\alpha\mu})\\
&+\frac{1}{180}\eta_{\mu\nu}t_{0}^{\alpha\beta}\Box
t_{0\alpha\beta}+\frac{1}{360}t_{0}^{\alpha\beta}\partial_{\mu}\partial_{\nu}t_{0\alpha\beta}
-\frac{1}{120}t_{0}^{\alpha\beta}(\partial_{\mu}\partial_{\alpha}t_{0\beta\nu}+\partial_{\nu}\partial_{\alpha}t_{0\beta\mu})\\
&+\frac{1}{60}t_{0}^{\alpha\beta}\partial_{\alpha}\partial_{\beta}t_{0\mu\nu}-\frac{1}{180}\partial_{\mu}t_{0}^{\alpha\beta}\partial_{\nu}t_{0\alpha\beta}
+\frac{1}{720}\eta_{\mu\nu}\partial_{\alpha}t_{0}^{\beta\gamma}\partial^{\alpha}t_{0\beta\gamma}\\&+\frac{1}{120}(\partial_{\mu}t_{0}^{\alpha\beta}\partial_{\alpha}t_{0\beta\nu}+\partial_{\nu}t_{0}^{\alpha\beta}\partial_{\alpha}t_{0\beta\mu})
-\frac{1}{60}\partial_{\alpha}t_{0\mu}^{\phantom{\mu}\beta}\partial_{\beta}t_{0\nu}^{\alpha}]+.....
\end{split}
\end{equation}
More generally, the particular solution for $g_n$ is uniquely determined once we specify that it vanishes at the boundary like $-(1/12)\rho^{6}\Box t_{n-2}$. Then it follows that it is independent of $t_n$ and doesn't contribute to the stress tensor at the n th order.

Now the particular solution at every order in the derivative expansion should by itself satisfy the scalar constraint. Let us see it explicitly for the particular solution for $g_2$. The particular solution chosen to vanish at the boundary like $-(1/12)\rho^{6}\Box t_{0}$ has an expansion of the above form (42). So by this choice, the coefficients of the Taylor expansion (now fixed by the source) will automatically agree with the general formulae, like those in (\ref{coeff}). These general formulae are automatically consistent with the scalar constraint. The scalar constraint also will be a linear differential equation for $g_n$ with a source term. The source term again is a (nonlinear) function of the corrections to the metric upto n-1 th order in the derivative expansion. The particular solution by itself will satisfy this equation. So the homogenous solution of the tensor equation for $g_n$ must also be a homogenous solution of the scalar constraint.

The homogenous solution of the tensor equation for $g_n$ which will be consistent with the scalar constraint is simply $-2b^4 f(\rho)t_{n\mu\nu}$, with $f(\rho)$ being given by (\ref{f}) and $t_{n\mu\nu}$ being an arbitrarily chosen correction to the hydrodynamic stress tensor involving n derivatives of the field theory coordinates z. However $t_{n\mu\nu}$ must be traceless and also satisfy the Landau gauge condition. Let us illustrate again by explicitly doing the Taylor series expansion of the homogenous solution to $g_2$ which is $-2b^4 f(\rho)t_{n\mu\nu}$. The Taylor expansion is as below:
\begin{equation}
g_{2\mu\nu}=t_{2\mu\nu}(\rho^4 + \frac{\rho^8}{4b^4}+\frac{\rho^{12}}{48b^8}+...) 
\end{equation}
Using the tracelessness and Landau gauge condition for $t_2$, one can check from the general formulae like those in (\ref{coeff}) that this is just the part of the metric determined by $t_2$ at the second order. Hence this should be the only homogenous solution that is consistent with the scalar constraint. Similarly at each order one can see that the part of the solution for $g_n$ which contains $t_n$ is proportional to $t_n$ and since the particular solution by choice contains all other terms, the homogenous solution should be always proportional to $t_n$. Then the tensor equation fixes the radial part of the homogeneous solution so that it should be $-2b^4 f(\rho)t_{n\mu\nu}$.

The vector constraint, at the n-th order in the derivative expansion, as we have argued before simply implies the conservation of the stress tensor upto n-1 th order. 

To summarize, these are the features of the derivative expansion in the Fefferman Graham coordinates.
\begin{itemize}
\item {At every order in the derivative expansion, the tensor equation for $g_n$ is a linear differential equation of the form of (\ref{tef}) involving derivatives in the radial coordinate only. The operators $D_1$, $D_2$, etc are the same at every order, while the source term $s_n$ is a nonlinear function of the various corrections to the metric upto n-1 th order.}
\item{The particular solution to the tensor equation for $g_n$ can be chosen to vanish at the boundary like $-(1/12)\rho^{6}\Box t_{n-2}$. With this choice the particular solution automatically satisfies the scalar constraint.}
\item{The homogenous solution to the tensor equation which is consistent with the scalar constraint is $-2b^4 f(\rho)t_{n\mu\nu}$ at very order, with f being given by (\ref{f}) and $t_{n\mu\nu}$ being an arbitrary n th order correction to the stress tensor which satisfy the tracelessness and the Landau gauge condition conditions.}
\item{The vector constraint at the n-th order just implies the conservation of n-1 th order stress tensor.}
\item{We can keep manifest Lorentz covariance at each order in the derivative expansion.}
\item{We can construct a solution corresponding to an arbitrary stress tensor because the homogenous solution of the tensor equation for $g_n$ at the n-th order is simply proportional to an arbitrarily chosen n-th order correction to the stress tensor.}
\it{At every order in the derivative expansion for any choice of the hydrodynamic stress tensor, the solution has finite radius of convergence away from the boundary, so all hydrodynamic stress tensors preserve the asymptotic AdS boundary condition.}
\end{itemize}

\section{Getting rid of naked singularities}
The comparative advantage of solving Einstein's equation of pure gravity in Fefferman Graham coordinates in the derivative expansion over doing the same in Eddington-Finkelstein coordinate system is that the constraints simplify dramatically and also we do not need to split the terms into tensors, vectors and scalars of SO(3), thus preserving manifest Lorentz covariance. The comparative disadvantage of the Fefferman-Graham coordinate system is that the regularity analysis is not straightforward. At the first order in the derivative expansion, the metric in Fefferman-Graham coordinates (\ref{mfo}) has a singularity at $\rho = \sqrt{2} b$. This is the location of the horizon at the zeroth order and the zeroth order metric itself is not invertible here. 

The first order perturbation has a \textit{log} piece which also blows up here. This singularity could be just a coordinate singularity in which case it could be removed by going to a different coordinate system as it happened for the boosted black brane, or it could be a real singularity. If it is a real singularity, it is naked because it coincides with the original horizon at late time. At late times the solution approaches a boosted black brane but since the horizon coincides with a real singularity, no infalling observer can continue life after reaching the horizon.

To analyse the singularity in the Fefferman-Graham coordinates we will simply translate the metric to Eddington-Finkelstein coordinates $(r,x)$. It will be of course suffice to change our coordinates near $\rho=\sqrt{2}b$, however, for the sake of completeness and better general understanding we will do the change of coordinates exactly upto first order in the derivative expansion. The Eddington-Finkelstein metric which we will get as a result of this translation will also be an exact solution of Einstein's equation upto first order in $x$-derivatives. We now return to the equations (\ref{trans}) in section 3 which gives the translation between the two coordinate systems. We still treat the Fefferman-Graham coordinates ($\rho(x,r)$, $z^{\mu}(x,r)$) as unknowns, but the third unknown is now the $G_{\mu\nu}(x,r)$ which appears in the Eddington-Finkelstein metric (\ref{EF}). The zeroth order solutions to these three are known and are given in (\ref{bbm}) and (\ref{bb}). To find the corrected solutions due to change in the Fefferman-Graham metric at first order it is straightforward to perturb these equations and solve them exactly at first order. The complete solutions to the three unknowns exact upto first order are:
\begin{equation}
\rho = \frac{\sqrt{2}b}{\sqrt{b^2 r^2 + \sqrt{b^4 r^4 -1}}}(1+bk(br)\frac{\partial . u}{3})\\\nonumber
\end{equation}
\begin{equation}\label{coc}
z^{\mu} = x^{\mu} + u^{\mu}bk(br) + u^{\mu}\frac{\partial . u}{3} b^2 k_{A}(br) + (u.\partial)u^{\mu}b^2k_{B}(br)
\end{equation}
\begin{equation}
\begin{split}
G_{\mu\nu} = &r^2 P_{\mu\nu} + (-r^2 +\frac{1}{b^4 r^2})u_{\mu}u_{\nu} + 2r^2 bF(br)\sigma_{\mu\nu} -r((u.\partial)(u_{\mu}u_{\nu})-\frac{2}{3}u_{\mu}u_{\nu}(\partial .u))\\\nonumber
&+\frac{(\gamma -1)b}{4}r^2 log(1-\frac{1}{b^4 r^4})\sigma_{\mu\nu}
\end{split}
\end{equation}
where,
\begin{eqnarray}
k(x) = \frac{1}{4}(log(\frac{x+1}{x-1}) - 2 arctan(x) + \pi)\\\nonumber
F(x) = \frac{1}{4}(log(\frac{(x+1)^2(x^2+1)}{x^4}) - 2 arctan(x) + \pi)
\end{eqnarray}
and $k_{A}(x), k_{B}(x)$ satisfy the following differential equations
\begin{eqnarray}
\frac{dk_A}{dx} = -\frac{x^2}{x^4-1}(k(x)+\frac{x}{\sqrt{x^4 -1}})\\\nonumber
\frac{dk_B}{dx} = \frac{1}{x\sqrt{x^4 -1}}-\frac{k(x)x^2}{x^4 - 1}
\end{eqnarray}
with the boundary condition that they vanish at $x=\infty$. One may easily check that if we do the Taylor series expansion of $\rho, z^{\mu}$ in 1/r, we can reproduce the results (\ref{sol}) of section 3 in which we have solved these equations using a power series ansatz.

The crucial point, as realized by authors of \cite{0712.2456} is that in the Eddington-Finkelstein coordinates if there is a blow-up in $G_{\mu\nu}(x,r)$ it should be a real singularity. For a general conformal fluid at first order with $\eta/s =\gamma/4\pi$, the corresponding solution in Eddington-Finkelstein coordinates has $G_{\mu\nu}(x,r)$ given by (\ref{coc}). Except for the \textit{log} term which appears in the last line, all other terms are well behaved for $r>0$ and the $log$ term blows up at $r=1/b$, the location of the unperturbed black brane horizon.  Only when $\gamma =1$, the coefficient of the \textit{log} term vanishes and so the naked singularity at $r=1/b$ is absent. For this value of $\gamma$ we have in fact reproduced the $G_{\mu\nu}$ of the Eddington-Finkelstein metric given by the authors of \cite{0712.2456}.

We learn the following general facts. The translation to Eddington-Finkelstein coordinates exists for an arbitrary solution in the Fefferman-Graham coordinates irrespective of whether there is any naked singularity or not. Also the Fefferman-Graham coordinates have a power series expansion in terms of the inverse of the radial Eddington-Finkelstein coordinates for all cases. For all cases, the change of coordinates also become singular at the location of the original horizon in the Eddington-Finkelstein coordinates which is $r=1/b$.

We can continue the regularity analysis to higher orders in the derivative expansion by solving the equations (\ref{trans}) for translating the solution from the Fefferman-Graham coordinates to Eddington-Finkelstein coordinates order by order in the derivative expansion as well. In this way at each order we will be able to determine what values the coefficients in the terms of the hydrodynamic stress tensor should have so that a naked singularity is avoided. It would be interesting to see if we can understand the values of these coefficients of the hydrodynamic stress tensor, more directly in terms of the geometry of the unperturbed boosted black brane horizon. 

We will conclude this section by emphasizing certain points.
\begin{itemize}
\item{We can think of translating to outgoing Eddington-Finkelstein coordinates also as an attempt to remove the singularity and then as expected the situation will be time-reversed. We will now need $\gamma = -1$ for regularity. In the boundary theory, all fluid dynamical solutions will then be time-reversed and our gravity solutions will be perturbed white-hole solutions exact upto first order in the derivative expansion.}

\item{We could have attempted to fix $\gamma$ by studying regularity at the horizon by computing curvature invariants (like $R_{\mu\nu\rho\sigma}R^{\mu\nu\rho\sigma}$). However, we do not know, if for these ``hydrodynamic'' space-times, checking that a finite number of curvature invariants do not blow up at the horizon will suffice to demonstrate regularity. So the best strategy is to translate to a coordinate system where the solution is explicitly regular upto first order in the derivative expansion and this is what we have done here. For the sake of completeness, however, we have studied a few curvature invariants and have found that the leading singularity of $R_{\mu\nu\rho\sigma}R^{\mu\nu\rho\sigma}$ at second derivative order vanishes for the right choices of $\gamma$ which are 1 and -1, the details of which are presented in Appendix B.}

\item{The derivative expansion in Fefferman-Graham coordinates is equivalent to the same in Eddington-Finkelstein coordinates to all orders in the derivative expansion even when the solutions do not have a regular horizon. This is so because the equations (\ref{trans}) for translating Fefferman-Graham coordinates to Eddington-Finkelstein coordinates can always be solved order by order in the derivative expansion as well. In fact, this is natural, because any asymptotic AdS solution can be written in the Fefferman Graham coordinates.}
 
\end{itemize}

\section{Discussion}
We will point out some implications of our results for dynamics in the universal sector of CFT. Our first result is that a solution of pure classical gravity is uniquely specified by the stress tensor. This implies that the dynamics of all states in the universal sector of the dual CFT at strong coupling and large N is completely determined by the conservation of the traceless stress tensor. The implication for dynamics on the CFT side is even more surprising than the result for classical theory of gravity itself. It is surprising because to characterise a state uniquely we would typically need the expectation values of infinite number of operators. However, it is not hard to give an example of a special sector of states with this property in a 2D CFT. These special states are spanned by $L_n |VAC>$ ($n\geqslant 2$)and are created by descendendants of the identity operator ($L_n$),with $n\geqslant 2$, acting on the vacuum. Each such state is uniquely character by the $L_0$ eigenvalue $n$, hence by the expectation value of the stress tensor T(z). Moreover each state $L_n |VAC>$ ($n\geqslant 2$) being an eigenvector of the Hamiltonian, the sector spanned by these states is closed under time evolution. It would be interesting to find such examples of class of states in CFTs in higher dimensions where the expectation value of the stress tensor uniquely identifies each member and moreover is closed under time evolution. The real question, however is, whether we can give an intrinsic microscopic description of the universal sector of CFTs with gravity duals. If we can achieve this, we will be able to understand better how the vev of the stress tensor and its conservation alone determines the dynamics in the universal sector completely.

Our second set of results are (a) all hydrodynamic stress tensors are free of the ``abcd'' type of pathology, which means that they preserve the asymptotic AdS boundary condition and (b) there is a unique hydrodynamic stress tensor for which there is no naked singularity. This means that the late time equilibriation in the boundary CFT can be determined by a unique and universal hydrodynamic stress tensor. The coefficients of the terms should be set to values which avoids formation of naked singularities in the bulk. It would be interesting to find out an intrinsic microscopic definition for the higher order coefficients of the hydrodynamic stress tensor, in terms of say, multi point correlations of the stress tensor. The first order coefficient, namely the viscosity has indeed such a definition in terms of two-point correlation function of the stress tensor and the validity of the definition can be verified by the AdS/CFT correspondence as well. So we may hope that a pure gravity analysis should suffice to arrive at similar definitions for the higher order coefficients in the hydrodynamic stress tensor. 

We would like to mention that while we were updating our work, it was a great pleasure to find out that our method has been generalised in \cite{0901.1487} to compute the stress tensor in the universal hydrodynamic sector of strongly coupled large N dual theories of various p-branes, which are in most cases non-conformal. We would like to take this opportunity to mention that since our method keeps the asymptotic boundary condition manifest, it could be given a preference whenever implementing the asymptotic boundary condition in Eddington-Finkelstein coordinates becomes laborious or complicated.

Finally, we would like to point out, that it will be interesting to find a physical understanding of the ``abcd'' type of pathology. Our results in the hydrodynamic sector gives support for claiming that whenever we have late-time equilibriation in the boundary stress tensor, this pathogy is absent. It will be interesting to find a real example with such a pathology and trace its physical origin. 

{\bf Acknowledgments:} We would like to thank Shiraz Minwalla, Romuald A. Janik, Rajesh Gopakumar, Ashoke Sen, Rukmini Dey, Sayantani Bhattacharya, Jyotirmoy Bhattacharya, Nabamita Banerjee and R. Loganayagam for various useful discussions which stimulated our work. We would like to thank Rajesh Gopakumar, in particular, for taking interest in our work from the very beginning and also for scrutinising the first draft. We would also like to thank TIFR and ICTS for inviting us to the Monsoon Workshop on String Theory where we got the opportunity to discuss our work with those we have mentioned before. We would also like to mention that we have used cluster computing facilities at Harish-Chandra Research Institute and the Mathematica based ``mathtensor'' package developed by M. Headrick for most of the computations presented in Appendix B. Finally we would like to thank the people of India for generously supporting research in fundamental physics.

\section*{Appendix A : Proof of the power series solution for $AdS_5$ asymptotics}

Here we will prove that any asymptotically $AdS_{5}$ solution of Einstein's equation with a negative cosmological constant, in the Fefferman-Graham coordinates, has a solution for $g_{\mu\nu}$ which is a power series in the radial coordinate when the boundary metric is flat. Though not explicitly mentioned in most of what follows, it should be kept in mind that here we are specifically investigating five-dimensional solutions with a flat boundary metric. At the end, we will mention if our proof can be generalised to other cases.

To simplify the proof we first rearrange the tensor and the scalar components of Einstein's equation (\ref{eom}) while keeping the vector components of Einstein's equation unchanged. The old scalar equation is added with an appropriate linear combination of the trace of the old tensor equation so that now it does not contain any term which has second derivative of $g_{\mu\nu}$ with respect to the radial coordinate $\rho$. Since the vector equation also does not contain any term with second derivative of $g_{\mu\nu}$ with respect to the radial coordinate we can now think of the vector and scalar components as a set of five constraint equations. We also change the tensor components of Einstein's equation by appropriately replacing $Tr(g^{-1}g^{\prime})$ using the new scalar equation. We do this so that now the tensor equation by itself is sufficient to determine all the $\rho^n$ coefficients of $g_{\mu\nu}$. The old tensor equation had the feature that to determine $g_{(8)\mu\nu}$, the coefficient of $\rho^8$ in $g_{\mu\nu}$, we had to use the scalar equation as well, but now this can be fully determined using the tensor equation alone. So our equations now are as below. 
\begin{equation}
\begin{split}\label{tensor}
&\frac{1}{2}g^{\prime\prime}-\frac{3}{2\rho}g^{\prime}-\frac{1}{2}g^{\prime}g^{-1}g^{\prime}+\frac{1}{4}Tr(g^{-1}g^{\prime})g^{\prime}-Ric(g)\\&+g[\frac{1}{6}R(g)+\frac{1}{24}Tr(g^{-1}g^{\prime}g^{-1}g^{\prime})-\frac{1}{24}(Tr(g^{-1}g^{\prime}))^2]=0
\end{split}
\end{equation}
\begin{equation}\label{vector}
\nabla_{\mu}Tr(g^{-1}g^{\prime})-\nabla^{\nu}g^{\prime}_{\mu\nu}=0
\end{equation}
\begin{equation}\label{scalar}
R(g)+\frac{3}{\rho}Tr(g^{-1}g^{\prime})+\frac{1}{4}Tr(g^{-1}g^{\prime}g^{-1}g^{\prime})-\frac{1}{4}[Tr(g^{-1}g^{\prime})]^2=0
\end{equation}
It is not difficult to see that we can use a power series ansatz to solve the tensor equation as at the n-th order. At the n-th order the only terms which can contain $g_{(n)_{\mu\nu}}$ or $Tr(g_{(n)})\eta_{\mu\nu}$ are $g^{\prime\prime}_{\mu\nu}$, $g^{\prime}_{\mu\nu}$ and $Tr(g^{-1}g^{\prime})g_{\mu\nu}$. Now since the tensor equation contains no term with $Tr(g^{-1}g^{\prime})g_{\mu\nu}$, at the n-th order,for $n>4$, the tensor equation gives us $n(n-4)g_{(n){\mu\nu}}/2 = f_{(n)\mu\nu}(t_{\rho\sigma})$, where $f_{(n)\mu\nu}(t_{\rho\sigma})$ is a polynomial in $t_{\rho\sigma}$ and its various derivatives with respect to the boundary coordinates only. Hence, for $n>4$, we can always solve $g_{(n){\mu\nu}}$ using the tensor equation alone.

We have now got to show that the power series we have so obtained as a solution to the tensor equation is consistent with the vector and scalar constraints. We will do this by the method of induction iterating over the various coefficients of $\rho^n$ in $g_{\mu\nu}$, order by order in n. We will first establish the following fact that the $\rho$-derivative of the vector and scalar constraints vanish when the tensor equation along with the vector and scalar constraints are satisfied. This just articulates the intuition that once the initial data consisting of $g_{\mu\nu}$ and $g_{\mu\nu}^{\prime}$ satisfy the vector and scalar constraints on hypersurface with a fixed value of the radial coordinate $\rho$, the dynamical evolution in $\rho$ should be such that the constraints should be automatically satisfied for any other hypersurface. To show this we will need the following:  
\begin{eqnarray}
{\Gamma^{\mu}_{\nu\sigma}}^{\prime}=\frac{1}{2}g^{\mu\alpha}(\nabla_{\nu}g^{\prime}_{\alpha\sigma}+\nabla_{\sigma}g^{\prime}_{\alpha\nu}-\nabla_{\alpha}g^{\prime}_{\nu\sigma})\\\nonumber
{R^{\mu}_{\nu\alpha\beta}}^{\prime}=\frac{1}{2}g^{\mu\gamma}[\nabla_{\alpha}\nabla_{\nu}g^{\prime}_{\gamma\beta}-\nabla_{\alpha}\nabla_{\gamma}g^{\prime}_{\nu\beta}-\nabla_{\beta}\nabla_{\nu}g^{\prime}_{\gamma\alpha}+\nabla_{\beta}\nabla_{\gamma}g^{\prime}_{\nu\alpha}]
\end{eqnarray}
One can use the tensor (\ref{tensor}) and scalar (\ref{scalar}) equations to write
\begin{equation}\label{bianchi}
\begin{split}
&R^{\mu}_{\nu}-\frac{1}{2}\delta^{\mu}_{\nu}R=\frac{1}{2}g^{\mu\alpha}g^{\prime\prime}_{\alpha\nu}-\frac{3}{2\rho}g^{\mu\alpha}g^{\prime}_{\alpha\nu}-\frac{1}{2}g^{\mu\alpha}g^{\prime}_{\alpha\beta}g^{\beta\gamma}g^{\prime}_{\gamma\nu}+\frac{1}{4}Tr(g^{-1}g^{\prime})g^{\mu\alpha}g^{\prime}_{\alpha\nu}\\&+\frac{5}{4\rho}Tr(g^{-1}g^{\prime})\delta^{\mu}_{\nu}-\frac{1}{4}\delta^{\mu}_{\nu}[Tr(g^{-1}g^{\prime\prime})-Tr(g^{-1}g^{\prime}g^{-1}g^{\prime})+\frac{1}{2}(Tr(g^{-1}g^{\prime}))^2]
\end{split}
\end{equation}
Now when all the equations (\ref{tensor}), (\ref{vector}) and (\ref{scalar}) are satisfied, the $\rho$-derivative of the vector constraint can also be written as:
\begin{equation}\label{vprime}
\begin{split}
(\nabla_{\mu}Tr(g^{-1}g^{\prime})-\nabla^{\nu}g^{\prime}_{\mu\nu})^{\prime}=&\partial_{\mu}[Tr(g^{-1}g^{\prime\prime}-\frac{3}{4}g^{-1}g^{\prime}g^{-1}g^{\prime})+\frac{1}{4}(Tr(g^{-1}g^{\prime}))^2]\\&-\nabla_{\nu}(g^{\alpha\nu}g^{\prime\prime}_{\mu\alpha}-g^{\alpha\beta}g^{\prime}_{\beta\gamma}g^{\gamma\nu}g^{\prime}_{\alpha\mu}+\frac{1}{2}g^{\nu\alpha}g^{\prime}_{\alpha\mu}Tr(g^{-1}g^{\prime}))
\end{split}
\end{equation}
Now comparing the right hand sides of (\ref{bianchi}) and (\ref{vprime}) using all the equations of motion again, we see that 
\begin{equation}
(\nabla_{\mu}Tr(g^{-1}g^{\prime})-\nabla^{\nu}g^{\prime}_{\mu\nu})^{\prime}=\nabla_{\nu}(R^{\nu}_{\mu}-\frac{1}{2}\delta^{\nu}_{\mu} R)
\end{equation}
So the Bianchi identity implies that the $\rho$-derivative of the vector equation should vanish when all the equations of motion are satisfied. We will now get to the scalar equation.

When the vector equation of motion (\ref{vector}) is satisfied we get
\begin{equation}
{R_{\mu\nu}}^{\prime}=-\frac{1}{2}R_{\alpha\mu}(g^{\alpha\beta}g^{\prime}_{\beta\nu})+\frac{1}{2}R^{\gamma}_{\nu\alpha\mu}(g^{\alpha\beta}g_{\beta\gamma}^{\prime})+\frac{1}{2}\nabla_{\mu}\nabla_{\nu}{Tr(g^{-1}g^{\prime})}-\frac{1}{2}\nabla^2g^{\prime}_{\mu\nu}
\end{equation}
This implies that when the vector equation of motion is satisfied, we have:
\begin{equation}\label{rprime}
R^{\prime}=-g^{\mu\nu}g^{\prime}_{\nu\sigma}g^{\sigma\alpha}R_{\mu\alpha}
\end{equation}
On the other hand the vanishing of the $\rho$-derivative of the scalar constraint (\ref{scalar}) ought to give us:
\begin{equation}\label{sprime}
\begin{split}
R^{\prime}=&-\frac{1}{2}Tr(g^{-1}g^{\prime}g^{-1}g^{\prime\prime})+\frac{3}{2\rho}Tr(g^{-1}g^{\prime}g^{-1}g^{\prime})\\&+\frac{1}{2}Tr(g^{-1}g^{\prime}g^{-1}g^{\prime}g^{-1}g^{\prime})-\frac{1}{4}Tr(g^{-1}g^{\prime})Tr(g^{-1}g^{\prime}g^{-1}g^{\prime})+\frac{1}{2\rho}[Tr(g^{-1}g^{\prime})]^2
\end{split}
\end{equation}
Now using the tensor and scalar equations of motion, we can see that the right hand sides of (\ref{rprime}) and (\ref{sprime}) are the same, or in other words the $\rho$-derivative of the scalar constraint indeed vanishes when all the equations of motion are satisfied. So we have established that the $\rho$-derivatives of the all the five constraints vanish when all the equations of motion are satisfied, or to state compactly 
\begin{equation}\label{induction}
(\ref{tensor}),(\ref{vector}),(\ref{scalar}) \Rightarrow (\ref{vector})^{\prime},(\ref{scalar})^{\prime}
\end{equation}
To prove that the power series solution of the tensor equation is consistent with the constraints, we will use the above at $\rho=0$. To obtain a condition for $g_{(n)\mu\nu}$ (the coefficient of $\rho^n$ in $g_{\mu\nu}$) from the tensor equation we need to differentiate it n-2 times with respect to $\rho$ and then set $\rho=0$. Similarly to obtain a condition for $g_{(n)\mu\nu}$ from the vector and scalar constraints we need to differentiate each of them n-1 times with respect to $\rho$ and then set $\rho=0$.

The vector and scalar constraints imply that $g_{(2)\mu\nu}$ should vanish while the tensor equation identically vanishes at this order. The tensor equation for $g_{(4)\mu\nu}$ which we have appropriately renamed $t_{\mu\nu}$, also identically vanishes while the vector constraint gives us the conservation equation $\partial^{\mu}t_{\mu\nu}=0$ and the scalar constraint gives the tracelessness condition $Tr(t)=0$. We can start our induction from here, since the three equations are all consistent with each other upto this order

Let us suppose, by the induction hypothesis that the solution for $g_{(n-1)\mu\nu}$ obtained from the tensor equation is consistent with the vector and scalar constraints. We now denote the m-th $\rho$-derivative as $m^{'}$. So, by induction hypothesis, the three equations $(n-3)^{\prime}(\ref{tensor})(\rho=0)$, $(n-2)^{\prime}(\ref{vector})(\rho=0)$ and $(n-2)^{\prime}(\ref{scalar})(\rho=0)$ are consistent with each other. Now we iterate by determining $g_{(n)\mu\nu}$ from the tensor equation, or in other words we solve
\begin{equation}
(n-2)^{\prime}(\ref{tensor})(\rho=0)
\end{equation}
But by induction hypothesis we can assume $(n-2)^{\prime}(\ref{scalar})(\rho=0)$ and $(n-2)^{\prime}(\ref{vector})(\rho=0)$ are consistent with the tensor equation. Now our result (\ref{induction}) for a general fixed $\rho$ hypersurface implies that
\begin{equation}\label{statement}
(n-2)^{\prime}(\ref{tensor}),(n-2)^{\prime}(\ref{vector}),(n-2)^{\prime}(\ref{scalar})\Rightarrow (n-1)^{\prime}(\ref{vector}),(n-1)^{\prime}(\ref{scalar})
\end{equation}
We can apply the above at $\rho=0$ \footnote{At $\rho=0$ the statement (\ref{statement}) has a non-trivial content strictly for $n>2$, because of the slight technicality that what we really need to use to find a condition for $g_{(n)\mu\nu}$ is that we need to differentiate $(\rho(\ref{scalar}))$ not really $(\ref{scalar})$ n-1 times. So at $\rho=0$, this result is trivial for the scalar constraint when $n=2$ and we do not need to use the result (\ref{statement}), but since the first step of induction starts from $n=4$, it is safe to use this in the iteration procedure.} to iterate and say that if the solution for $g_{(n-1)\mu\nu}$ from the tensor equation is consistent with the constraints so would the solution for $g_{(n)\mu\nu}$ from the tensor equation be. This completes the proof by induction that the power series solution of the tensor equation is consistent with the constraints.

Let us see if our proof can be generalised to other cases, in particularly for all dimensions if the boundary metric is flat. The only change in the equation of motion happens to be the coeffiecient of $g_{\mu\nu}^{\prime}$ in the tensor equation. Let us, for example, take the case when the number of boundary coordinates is six. We can check by hand that all $g_{(n)\mu\nu}$ vanish for all n such that $0<n<6$ and $g_{(6)\mu\nu}$ cannot be determined from the tensor equation for an exactly similar reason as for $g_{(4)\mu\nu}$ when the number of boundary coordinates was four, namely the tensor equation identically vanishes. The vector and scalar constraints imply conservation and tracelessness of $g_{(6)\mu\nu}$ implying that it should be identified with the stress tensor (and indeed it has been shown in \cite{0002230} that this agrees with with the Balasubramanian-Krauss stress tensor). We can begin our induction, from here as before and hence our proof generalises. So, the general problem in applying the induction is to show that the equations of motion are consistent with the power series ansatz at $g_{(d)\mu\nu}$. We have not been able to prove it generally but we have checked it upto $d=6$. The same problem appears when we try to apply induction to prove the validity of the power series solution when the number of boundary coordinates is odd, but the boundary metric is arbitrary. Before we apply induction, we need to prove that the power series works at $g_{(d)\mu\nu}$, (in fact this is harder to show, because when the boundary metric is not flat $g_{(n)\mu\nu}$'s do not vanish for $0<n<d$). However, Fefferman and Graham have proved the validity of the power series solution by a different method for an arbitrary boundary metric when the number of boundary coordinates is odd.

\section*{Appendix B: On fixing $\eta/s$ by calculating curvature invariants}

We have already done the regularity analysis of our first order solution in Fefferman Graham coordinates by translating to Eddington-Finkelstein coordinates where the regularity or irregularity becomes manifest. However, one may ask if the regularity analysis can be done also by calculating some curvature invariants. We will see that indeed at the first order, this analysis can also be done by calculating an appropriate curvature invariant, but we will argue that there may not be a finite number of curvature invariants which can be reliably used to fix all the coefficients in the hydrodynamic stress tensor at higher orders in the derivative expansion.

Before we do that, we want to point out that though the metric in Fefferman-Graham coordinates and in Eddington-Finkelstein coordinates could be made coordinate equivalent upto any given order in the derivative expansion for an arbitrary hydrodynamic stress tensor, the curvature invariants calculated from the two metrics will typically never be the same! Let us examine why this should happen at the first order itself. Any typical curvature invariant, like the Ricci scalar R itself, will show a divergence only when we expand it to second order in derivatives of the boundary coordinates. In this case, this should be so, because the metric in either coordinate system is a solution of the equations of motion upto first order in derivatives of boundary coordinates. However, the second order piece in R calculated from the metric in either coordinate system will not be the same, because the two metrics are related by a coordinate transformation only upto first order in derivatives. In fact we will explicitly demonstrate that R itself can be used to fix the value of $4\pi\eta/s$ in the Eddington-Finkelstein metric at first order but not in the Fefferman-Graham metric at first order. So the procedures of using curvature invariants to fix the coefficients in the hydrodynamic stress tensor in the two coordinate systems are indeed very different!

Another crucial aspect should be kept in mind because this also features in comparing curvature invariants calculated from the metrics in the two coordinate systems. Fundamentally, solving Einstein's equations in either of the two coordinate systems involves a trade-off between manifest regularity and manifest asymptotic boundary condition. The solution in Eddington-Finkelstein coordinate system at the zeroth order and also at the first order for the right value of $4\pi\eta/s$ are manifestly regular so any curvature invariant calculated at the horizon will be regular to all orders as well. However, the solution preserves the asymptotic AdS boundary condition only upto first order in derivatives as it can be translated to Fefferman-Graham coordinate system only upto that order. The solution in Fefferman-Graham coordinate system at first order, of course preserves boundary condition to all orders, but even for the right choices of $4\pi\eta/s$ it is not regular to all orders. In other words, for the right choice of $4\pi\eta/s$ all order divergences should vanish when we calculate curvature invariants from the metric in Eddington-Finkelstein coordinate system, but in case of the solution in Fefferman-Graham coordinates at first order, at most the leading divergence at the second order vanishes for the right choice of $4\pi\eta/s$. In fact, for certain curvature invariants even that do not happen. Of course, eventually if we add a right second order correction to the Fefferman-Graham metric, all divergences in the curvature invariants at the second order should vanish, but still divergences at higher orders will remain and so on. We will illustrate the first order case with examples below.

To compute curvature invariants it is useful to first choose a velocity and temperature profile. As mentioned before, the vector constraint in Einstein's equations of motion demand that the velocity-temperature profile should be a solution of the relativistic Euler equation 
\begin{equation}\label{euler}
\frac{\partial_{\mu}b}{b} = (u.\partial)u_{\mu} - u_{\mu}\frac{\partial.u}{3}  
\end{equation}
We call our boundary coordinates (t, x, y, z) and we select the following static velocity profile which is a relativistic version of laminar flow
\begin{equation}
u^{\mu} = \frac{1}{\sqrt{1- a^2 y^2}}(1, ay, 0, 0) 
\end{equation}
where $a$ is a constant of dimension 1/length. The advantages of using this velocity profile are twofold, namely,
\begin{itemize}
\item {The relativistic Euler equation gives us that temperature, hence b, should be a constant.}
\item {It is easy to employ the derivative expansion by using the following trick. We note that the only non-trivial derivatives of the boudary coordinates are the y-derivatives. Any y-derivative of the velocities will bring in an extra $a$ which is unpaired with a $y$ so that it picks up the right dimension. Hence to do the  derivative expansion we may first set $y = p/a$ and simply do a Taylor expansion in $a$ about $a=0$. The correct dimensionless parameter of the derivative expansion, of course will be $ab$.}
\end{itemize}
We can use the above velocity-temperature profile in the first order solution in any coordinate system. Though away from the boundary the boundary coordinates (or, in other words, the field-theoretic coordinates) in a given coordinate system will mix with all the coordinates in another coordinate system, at the boundary they will always align with other. This is, how solutions in two different coordinate systems come to share the same boundary stress tensor and also the same conservation equation, which in this case, is the relativistic Euler equation.

If we use the above velocity-temperature profile to calculate R in the Eddington Finkelstein coordinate system we will find that
\begin{equation}
\begin{split}
&R = -20  + a^2 \frac{1}{8(1-a^2 y^2)^2 b^4 r^6}\\&[\frac{(\gamma-1)(b^2 r^2(9+3\gamma -2\pi)-16b^5 r^5 +2\pi b^6 r^6))}{(br-1)(1+br+b^2 r^2+b^3 r^3)}+ \\&(\gamma- 1)(\gamma+1-8b^3 r^3 )b^2 r^2Log(1-\frac{1}{b^4r^4}) + O(1)]+ O(a^3)
\end{split}
\end{equation}
At the zeroth order in $a$, R should of couser be -20 and at order $a$, R should of course vanish because our metric is a solution of equations of motion upto first order. At order $a^2$, we indeed expect some divergence at the horizon, which is at $r=1/b$, because the metric is explicitly not regular there unless $\gamma = 4\pi\eta/s =1$. We see that when $4\pi\eta/s =\gamma =1$ all divergences go away. This feature replicates also at higher orders in $a$. \footnote{We would like to thank Sayantani Bhattacharya for confirming that this indeed happens for arbitrary velocity and temperature profiles.} On the other hand, if we calculate R from the Fefferman-Graham metric at first order, we get
\begin{equation}
\begin{split}
&R = -20 +a^2[\frac{128b^{10}\rho^{8}(12b^{4}\gamma^2 +4b^2 \rho^2 + 3\gamma^2 \rho^4)}{(1-a^2 y^2)^2 (4b^4 -\rho^4 )^2 (4b^4 +\rho^4)^3}\\&
+\frac{16b^6 \rho^4 \gamma^2}{(1-a^2 y^2)^2 (4b^4 +\rho^4)^2}Log(\frac{4b^4 -\rho^4}{4b^4 + \rho^4})] + O(a^3 )
\end{split}
\end{equation}
At order $a^2$, we see that there is a leading inverse power two divergence for any value of $\gamma$ and a subleading log divergence except when $\gamma = 0$. So this is useless to figure out the right value of $\gamma$. Of course this will certainly be useful to fix certain coefficients of the hydrodynamic stress tensor at second order, because these divergences should go away for any right second order correction to the Fefferman-Graham metric.

It turns out, however, that, $R_{\mu\nu\rho\sigma}R^{\mu\nu\rho\sigma}$ can be used to fix the value of $\gamma$ in the Fefferman-Graham metric. We get
\begin{equation}
\begin{split}
&R_{\mu\nu\rho\sigma}R^{\mu\nu\rho\sigma} = \frac{4(1280b^{16}+1280b^{12}\rho^4 + 2784b^{8}\rho^{8}+80b^{4}\rho^{12}+5\rho^{16})}{(4b^4 +\rho^4 )^4}\\&
-a^2[\frac{2(1-\gamma^2)b^6}{(1-a^2 y^2)^2 (\rho - \sqrt{2}b)^4}+O(\frac{1}{(\rho -\sqrt{2}b)^2}) + O(Log(\sqrt{2}b-\rho)) + O(1)]
\end{split}
\end{equation}
We see that the zeroth order piece is always finite and independent of $\gamma$ and at order $a$ (for some reason we do not understand) the scalar vanishes. However, at order $a^2$, we find that when $\gamma$ is 1 or -1 the leading divergence at $\rho=\sqrt{2}b$ goes away, though, the subleading divergences remain and as before, they should disappear when we add any right second order contribution to the Fefferman-Graham metric. We are also not sure, if by computing $R_{\mu\nu\rho\sigma}R^{\mu\nu\rho\sigma}$ itself we can fix the values of all the coefficients in the hydrodynamic stress tensor at second order. To fix all the coefficients of the second order hydrodynamic stress tensor, one may have to look for another appropriate curvature invariant. 

It is certainly, worth exploring, if the ``hydrodynamic'' Fefferman-Graham solutions are ```special'' enough so that computing a finite number of curvature invariants will suffice to determine regularity, hence in fixing all the coefficients in the hydrodynamic stress tensor to all orders. We will leave this for a future work. Nevertheless, our procedure of fixing the coefficients in the hydrodynamic stress tensor by translating to Eddington-Finkelstein coordinate system works for all orders in the derivative expansion.

\end{document}